\documentclass[sigconf]{acmart}

\AtBeginDocument{%
}

\copyrightyear{2026}
\acmYear{2026}
\setcopyright{cc}
\setcctype{by}
\acmConference[ICMI '26]{INTERNATIONAL CONFERENCE ON MULTIMODAL INTERACTION}{October 05--09, 2026}{Napoli, Italy}
\acmBooktitle{INTERNATIONAL CONFERENCE ON MULTIMODAL INTERACTION (ICMI '26), October 05--09, 2026, Napoli, Italy}
\acmDOI{10.1145/3776574.3831180}
\acmISBN{979-8-4007-2318-6/2026/10}

\begin{document}

\title{Smiling Regulates Emotion During Traumatic Recollection}

\author{Marcus Ma}
\email{mjma@usc.edu}
\affiliation{%
  \institution{University of Southern California}
  \city{Los Angeles}
  \state{California}
  \country{USA}}
  
\author{Emily Zhou}
\affiliation{%
  \institution{University of Southern California}
  \city{Los Angeles}
  \state{California}
  \country{USA}}

\author{Leonard Ludwig}
\affiliation{%
  \institution{Freie Universität Berlin}
  \city{Berlin}
  \country{Germany}
}

\author{Julia Hörath}
\affiliation{%
  \institution{Freie Universität Berlin}
  \city{Berlin}
  \country{Germany}
}

\author{Christina Winkler}
\affiliation{%
  \institution{Freie Universität Berlin}
  \city{Berlin}
  \country{Germany}
}

\author{Kleanthis Avramidis}
\affiliation{%
  \institution{University of Southern California}
  \city{Los Angeles}
  \state{California}
  \country{USA}}
  
\author{Tiantian Feng}
\affiliation{%
  \institution{University of Southern California}
  \city{Los Angeles}
  \state{California}
  \country{USA}}
  
\author{Gabor Mihaly Toth}
\affiliation{%
  \institution{University of Luxembourg}
  \city{Esch-sur-Alzette}
  \country{Luxembourg}
}

\author{Alina Bothe}
\affiliation{%
  \institution{Freie Universität Berlin}
  \city{Berlin}
  \country{Germany}
}
  
\author{Shrikanth Narayanan}
\affiliation{%
  \institution{University of Southern California}
  \city{Los Angeles}
  \state{California}
  \country{USA}}

\renewcommand{\shortauthors}{Ma et al.}

\begin{abstract}
We study when, where, and why 978 Holocaust survivors smile in video testimonies. We create an automatic smile detection model from facial features with an F1 of 85\% and annotate detected smiles under two established taxonomies of smiling. We produce narrative features on 1,083,417 transcript sentences as well as emotional valence from three different modalities: audio, eye gaze, and text transcript. Smiling rates are associated with specific semantic topics, narrative structures, and temporal syntaxes across the corpus. Smiles often occur during periods of intense negative affect and we find negative-affect smiles are associated with more positive subsequent valence trajectories across all three modalities. Smiling reduces eye dynamics and blink rates, with both of these effects modulated by narrative valence. Taken together, we conclude that smiling plays a critical role in regulating emotion and social interaction during traumatic recollection.
\end{abstract}

\ccsdesc[300]{Applied computing~Psychology}
\ccsdesc[300]{Applied computing~Sociology}
\ccsdesc[500]{Human-centered computing}
\ccsdesc[500]{Computing methodologies~Discourse, dialogue and pragmatics}

\keywords{smiling, emotion, social interaction, traumatic recollection, multimodal interaction}

\maketitle


\section{Introduction}
\label{sec:intro}

Smiles are complex social behaviors. While we humans instinctually smile when we are happy \cite{freedman_1964}, we also smile during periods of great distress, as reactions to surprise, and to mask our underlying emotions. Smiling, therefore, is not simply a valence barometer; it is a carefully constructed display of social expression that balances internal affect and the emotions one wishes to display to the world. In this work, we investigate the social and narrative role that smiling plays during interviews of traumatic recollection with survivors of the Holocaust. We find this nonverbal behavior acts as a mediator of emotion, both in the emotions embedded into recalled narratives and into the social dynamics captured between survivor and interviewer.

Existing smile taxonomies classify smiles directly via internal affect \cite{ekman_1982} and social role \cite{martin_2017}, but we find in our investigations that the emotional weight and complexity of survivor testimonies renders these existing taxonomies overly simplistic. We conducted a pilot smile annotation study under these two taxonomies with trained Holocaust historians and find smiling is associated with both underlying and projected emotional expressions and with changes in narrative emphasis. From these results, we performed a large-scale automatic annotation of narrative features and investigate the complex multimodal interaction that occurs during smiling between facial movements, eye gaze, audio, and semantic content.

We identify associations between smiling and two types of survivor emotion: the ``narrative valence'' of the historical content being recollected and the ``present-day valence'' of the in-the-moment social dynamics when interfacing with interviewers. We find smiling typically occurs in emotionally congruent settings, such as during topics associated with pre-war and childhood events, but that almost a fifth of all smiles occur during moments of negative valence. Negative-valence smiles seem to be associated with emotional regulation, as the presence of such smiles improves the valence trajectories of surrounding sentences both in narratives and present-day, and across the three modalities of audio, eye gaze, and transcript. Smiling is also associated with differences in eye dynamics and blink rates that vary with narrative valence. Taken together, we assert that smiling plays a critical role in regulating emotion and social interaction during traumatic recollection.

\section{Dataset}
\label{sec:dataset}

The video data of this study were curated from the USC Shoah Foundation's Visual History Archive (VHA)\footnote{\url{https://vha.usc.edu/home}}, which contains over 55,000 video testimonies of survivors and witnesses of the Holocaust and other genocides recorded between primarily 1994 and 1999 in over 50 countries and in over 30 languages. For the present study, we limit our analysis to 978 English-speaking subjects who survived internment at the various concentration camps.

\paragraph{Interview contents}
We selected a subset of 978 English-language interviews with each interview split into an average 4.1 tapes of about 30 minutes, totaling 1,965 hours of playtime. Each interview was conducted in a similar format, with the survivor seated and still in their home facing an interviewer behind the camera. Each survivor gave testimonies on pre-war life, Holocaust experiences, and post-liberation life, for an average interview duration of two hours.

\paragraph{Complexity and weight of the interviews}
In these interviews, Holocaust survivors bear witness to a crime whose cruelty and
scale defy human comprehension. In describing their experiences of deportation, camps, extreme hunger, medical experiments on human beings, arbitrary killings, and mass extermination, the interviewees repeatedly reach the limits of what can be said and described \cite{amery_1980, langer_1995, laub_2016}, owing to the Holocaust's very nature as an irrational event that resists ``adequate verbalization'' \cite{quindeau_1995}. The narratives often consist of deeply traumatizing memories, of which the processing shaped survivors’ entire lives and survival after liberation in 1945. Whether and how experiences of the past can be remembered and narrated in the interview situation depends fundamentally on the interview setting itself and on the social - and sometimes, therapeutic - relationship that interviewer and interviewee enter into \cite{laub_2016}. Dori Laub, a Holocaust survivor and psychologist, describes video interviews as a ``testimonial intervention'' \cite{laub_2016}. At the same time, the conditions under which the interview takes place shape the respective testimony itself \cite{bothe_2019}.

\paragraph{Initial observations about smiling}
Smiling plays a central role during communication and fulfills a variety of functions \cite{martin_2017, kraut_1979, niedenthal_2010}, and during initial analysis, we often found incredibly complex emotional processing occurring simultaneously. We found smiles contained expressions of mixtures of pride, regret, and self-irony; unexpected smiles during extreme negative valence; and disassociation when smiling. While most subjects would smile during happy recollections, we found negative-valence smiling a very individually-variable behavior.

\paragraph{Additional multimodal data}
In addition to the interview footage, the dataset also includes the following, derived from prior works:

\begin{itemize}
\item \textbf{Original Video}: 320$\times$240 at 29.97 or 25\,fps (H.264) \cite{shoah}
\item \textbf{Original Audio}: 44.1\,kHz stereo (AAC) \cite{shoah}
\item \textbf{Annotated Transcript}: time-aligned human annotation and diarization \cite{shoah}
\item \textbf{Metadata keywords}: minute-level annotations from a closed set of Holocaust-specific keywords~\cite{shoah_2022}
\item \textbf{Facial Action Units}: 17 facial Action Units (AUs)~\cite{ekman_2002} from OpenFace~\cite{Baltrusaitis2018}, captured at 30\,fps
\item \textbf{Emotion from audio}: Valence, Arousal, and Dominance (VAD) \cite{mehrabian_1996} emotion values, temporally-aligned \cite{lertpetchpun_2025, feng_2025}
\item \textbf{Emotion from eyegaze}: temporally aligned values of VAD from five seconds of prior eye movement \cite{ma_2026}
\item \textbf{Localized eye movement}: subject eye movement that is normalized relative to inferred interviewer position \cite{ma_2026}
\end{itemize}

\begin{figure}[t]
  \centering
  \includegraphics[width=\columnwidth]{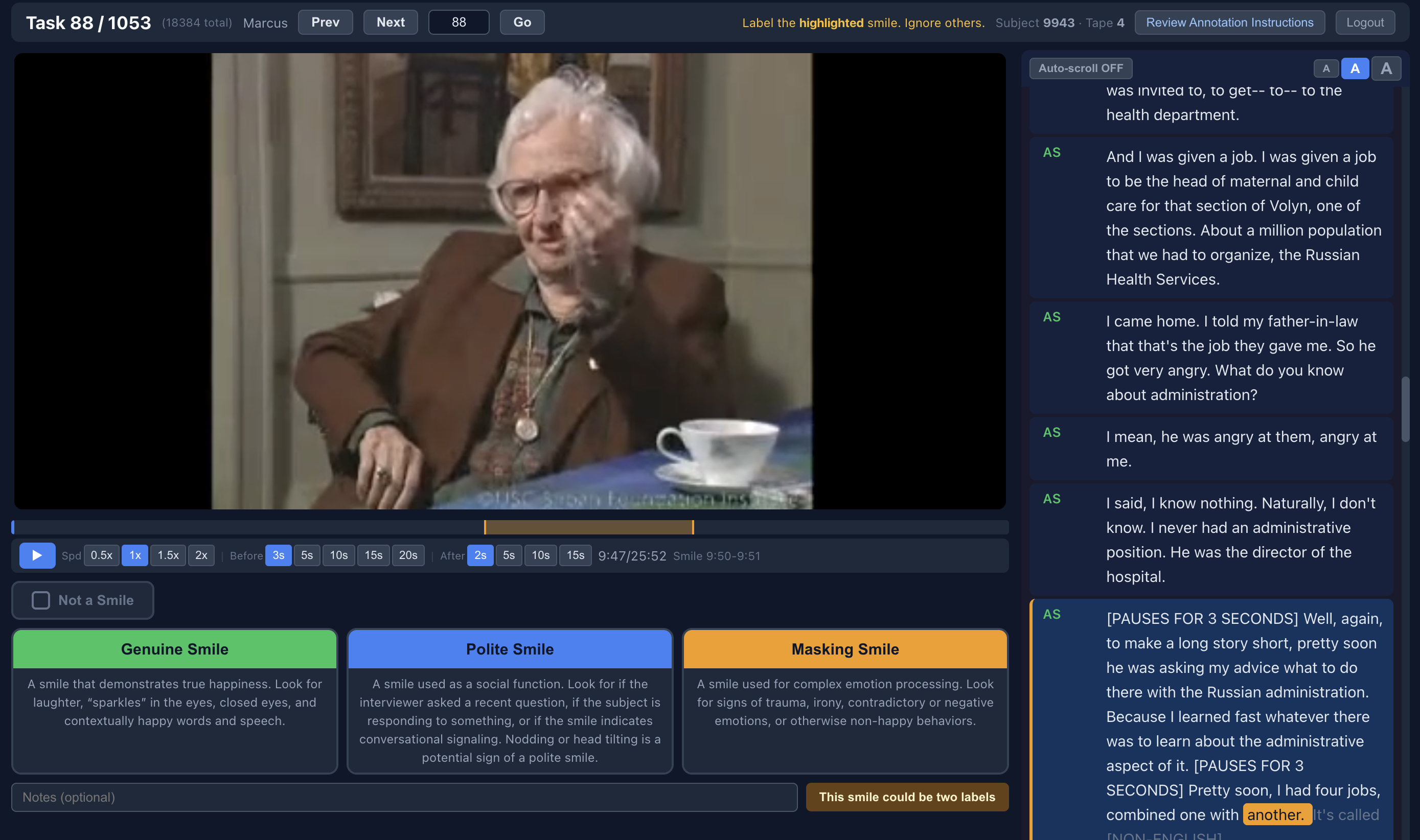}
  \caption{Annotation interface for the smiling label tasks.}
  \Description{Screenshot of a custom annotation interface showing interview video beside a time-aligned transcript for labeling smiles.}
  \label{fig:ui}
\end{figure}

\section{Related Work and Pilot Study}
\label{sec:pilot}

The modern scientific study of smiling began with the discovery of the ``Duchenne smile'' in 1862 \cite{duchenne_1862}, which contrasted smiles that engaged just the corners of the lip with those that also had raised cheeks and eyelid lowering. Duchenne smiles, which are associated with genuine signs of joy and positive valence \cite{ekman_1990}, have since been further formalized as specific Action Unit (AU) activations under \citeauthor{Ekman1978}'s Facial Action Coding System \cite{Ekman1978}.

\citeauthor{ekman_1982} proposed a three-class taxonomy of felt, false, and miserable smiles derived from affective state and physical markers \cite{ekman_1982}; in recent years, building on evidence that smiling evolved chiefly as a social tool \cite{fridlund_1994}, \citeauthor{niedenthal_2010, martin_2017} established another taxonomy with smiles serving three social roles of reward, affiliation, and dominance \cite{martin_2017}. In dialogue interactions specifically, smiling has long been known to be a direct communication device, where humans subconsciously time the onset of smiles to maximize communicative impact and social signaling \cite{chovil_1991, bavelas_2000}.

We begin our analysis with a pilot study where six annotators, including four trained Holocaust historians with graduate degrees, labeled smiles under the two taxonomies. Annotators met weekly over the course of two months to discuss disagreements, update instructions, and come to consensus on labels. Below, we describe the annotation process (\S\ref{subsec:pilot}), the agreement scores (\S\ref{subsec:low_agreement}), some analysis and results (\S\ref{subsec:pilot_results}), and a derived smile detector (\S\ref{subsec:smile_detection}).

\subsection{Pilot Annotation and Analysis}
\label{subsec:pilot}

\paragraph{Annotation tasks} Annotators labeled each smile under two taxonomies summarized in Table~\ref{tab:smile-taxonomies}: the \textit{Genuine/Polite/Masking} scheme adapted from \citet{martin_2017} that captures social functions of the smile, and the \textit{Felt/False/Miserable} scheme from \citet{ekman_1982} that characterizes the underlying emotional authenticity of the smile. In addition to assigning labels to smiles, annotators performed open coding \cite{strauss_1990} on randomly selected smiles by describing in plain text why they thought the smile was occurring without forcing a taxonomic label.

\begin{table}[t]
\centering
\caption{The two pilot smile taxonomies emphasizing social functions \cite{martin_2017} and emotional authenticity \cite{ekman_1982}.}
\label{tab:smile-taxonomies}
\small
\begin{tabular}{p{0.2\columnwidth} p{0.7\columnwidth}}
\toprule
\textbf{Label} & \textbf{Definition} \\
\midrule
\multicolumn{2}{l}{\hspace{3cm} \textbf{Social Function}} \\
Genuine  & Spontaneous smile reflecting felt positive affect \\
Polite   & Deliberate smile used socially for agreement\\
Masking  & Smile used to conceal negative emotion \\
\midrule
\multicolumn{2}{l}{\hspace{2.5cm}\textbf{Emotional Authenticity}} \\
Felt     & Authentic smile with smooth onset/offset \\
False    & Voluntary, unfeeling smile isolated to the lips \\
Miserable & Negative affect smile surrounding sad facial features \\
\bottomrule
\end{tabular}

\end{table}

While \citeauthor{martin_2017}'s original taxonomy was Reward / Affiliative / Dominant, we used the terms Genuine / Polite / Masking as survivors rarely exhibited dominance. We replaced Dominant with Masking as we commonly observed a type of smile that did not map cleanly onto \citeauthor{martin_2017}'s taxonomy, a kind of grimace meant to mask the horrible reality of what the interviewee was describing. This smile seemed to play both a social and emotionally regulatory role as it was associated with comforting the interviewer as well as personally processing emotion.

\paragraph{Interface} We built a custom annotation interface that displays the time-aligned transcript alongside the interview video and audio (Figure~\ref{fig:ui}). Each annotation task shows a video clip containing 3\,s before the detected smile, the entire smile segment, and 2\,s after, with the median smile lasting 1.37\,s. Annotators could re-watch smiles as needed and relied on facial features, voice, transcript, and social context and met weekly over two months to discuss which elements of the interview motivated their labels. 

\paragraph{Smile Candidate Collection}
\label{sec:smile_filtering}

To gather smile candidates, for each video, we analyzed framewise AU12 (lip corner puller) intensity from OpenFace's AU detector. Because of the low resolution of the archival footage, the OpenFace intensities are likely noisy estimates. We Gaussian-smoothed this time series with $\sigma=0.133\,s$, thresholded for AU values $> 1.0$\footnote{AU12's intensity ranges from 0 to 5 with the median frame from our data being 0.}, removed occurrences of less than half a second, and merged segments occurring within half a second. This process yielded 287,991 smile candidates across 3,649 videos.

In order to increase the likelihood of smiles during annotation, we applied additional filters during both taxonomy tasks: for the first we increased the AU12 threshold to $> 1.5$ and for the second we trained a logistic regression model on 17 AUs (including AU12) that best fit the F1 of the first set labels and set the regression threshold at 75\% specificity. Annotators had the ability to rate a stimulus as ``not a smile'' which we used later to train a smile detection model. We separately annotated a third set of 250 smiles whose ranges lay beyond the AU12 $>1.0$ filter but before either of the additional filters to have representative coverage of the entire candidate set.

\begin{table}[t]
\centering
\caption{Pilot annotation results. Fleiss' $\kappa$ reported for
         the 4-class (three smile labels + \textit{not a smile})
         and binary (smile vs.\ no smile) settings.}
\label{tab:annotation-results}
\small
\begin{tabular}{lrrr}
\toprule
\textbf{Taxonomy} & \textbf{Candidates} & \textbf{$\kappa$ (4-class)} & \textbf{$\kappa$ (binary)} \\
\midrule
Genuine / Polite / Masking & 795 & 0.244 & 0.575 \\
Felt / False / Miserable   & 452 & 0.455 & 0.714 \\
\bottomrule
\end{tabular}
\end{table}

\subsection{Annotator Agreement and Pilot Analysis}
\label{subsec:low_agreement}

We report annotation numbers and agreement scores for both the 4-class setting (the three labels plus ``not a smile'') and the binary smile vs. no smile setting in Table~\ref{tab:annotation-results}. Annotator agreement on both taxonomies was 0.244 and 0.455. We hypothesize since the first taxonomy emphasizes the social role of smiling while the second emphasizes the affective and physical role, annotators could more easily determine the latter as half the social dynamic between the subject and interviewer was hidden behind the camera. Agreement for the more objective task of smile vs. no smile detection was 0.575 and 0.714 across both taxonomies, and we attribute this improvement to the replacement of the AU12 $ > 1.5$ secondary filter with a 17 AU logistic regression model that best fit the first annotation set.

We note the difficulty of the annotation task and the subjectivity required. During weekly discussion meetings, it was not uncommon for a single label to be discussed for many minutes without achieving consensus. Low annotator agreement is not just noise but a reflection of the inherent subjectivity and difficulty in the data \cite{chochlakis_2025, aroyo_2015}; additionally, annotators frequently noted that the three-class smiling labels applied were often too broad to apply perfectly onto the annotations. Survivors frequently reflected on incredibly emotionally charged topics as death, slavery, and unimaginable cruelty and these conditions presented situations with complex and ultimately subjective decision making. For example, in one case, a former doctor smiled during recollection of digging up corpses to learn anatomy because the hospital ran out of cadavers; even after explicit discussion attempting to form consensus, two annotators could not agree if the smile was genuine or masking.

\begin{figure}[t]
  \centering
  \includegraphics[width=0.9\columnwidth]{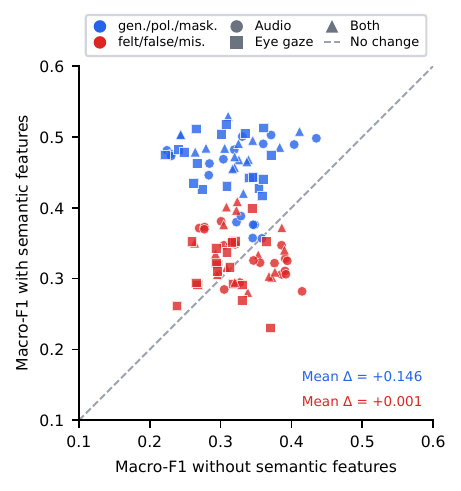}
  \caption{Macro-F1 with and without LLM transcript features for \citeauthor{martin_2017} taxonomy (blue, +0.146 avg) and \citeauthor{ekman_1982} taxonomy (red, +0.001 avg).}
  \Description{Bar chart comparing macro-F1 scores with and without LLM transcript features for two smile taxonomies.}
  \label{fig:text-ablation}
\end{figure}

\paragraph{Orientation within Holocaust literature} Survivors of the Holocaust, having survived by mere chance and confronted with the felt meaninglessness of their own existence, have been historically observed to process emotions in atypical ways known as survivor syndrome \cite{niedeland_1980}. In the testimonies, annotators remarked one manifestation of this disposition is often concealing emotion behind a smile. The impact of the Holocaust has been described as a ``rupture in civilization'' \cite{diner_2007} that has shaped subsequent generations around the world culturally, emotionally, and pervasively \cite{auerhahn_1984, quindeau_1995}. Given the severity of this atrocity, even the notion of objective scholarship must be called into question.

\paragraph{Open coding analysis} Interpretation of the open coding of 200 smiles reveals smiling often served both reflective and storytelling roles, with common patterns including fond memory recall, narrative emphasis, and irony. Consistent with existing literature on understanding how emotion shapes personal narrative, we found the most striking difference between the ``Narrated I'' (how the subject felt then) vs. the ``Narrating I'' (how the character feels now) \cite{lazarus_1991, bamberg_1997}. We observed associations between smiling and two kinds of emotion: the \textbf{``narrative valence,''} meaning the emotion the interviewee likely felt during the events of their stories, and the \textbf{``present-day valence,''} or what they felt speaking in the interview.

\subsection{Semantics as a Predictive Signal}
\label{subsec:pilot_results}
We attempted to isolate which non-facial features (meaning \textbf{not} using the AU features) best predicted smile class across both taxonomies. We trained 108 lightweight classifiers to predict the three labels over a large sweep of features and models: three architectures (MLP, GRU, CNN), three modality combinations (include/exclude audio emotion, eyegaze emotion, LLM-derived semantic features), and six temporal windows surrounding the smile with leave-one-out cross validation. To derive these features, we prompted the open-source LLM GPT-OSS-120B \cite{gpt-oss-120b} to assign 0 to 10 scores for ten prosodic features relating to affect and socialization:

\begin{quote}
    courtesy, nostalgia, humor, emotional weight, coping, direct interviewer response, topic shift, self-reference, positivity, negativity
\end{quote}

We find most models perform marginally better than chance, but that on the social roles taxonomy, inclusion of the LLM-derived features dramatically improves performance while no effect is recorded on the other taxonomy (Figure~\ref{fig:text-ablation}). We use this finding to motivate our large-scale collection of semantic narrative features in \S\ref{sec:open-coding}.

\subsection{Smile Detection Model}
\label{subsec:smile_detection}

Because 4-class agreement was low, we collapsed taxonomy labels into a binary ``smile'' vs.\ ``not a smile'' task for detector training.

\begin{figure}[t]
  \centering
  \includegraphics[width=0.85\columnwidth]{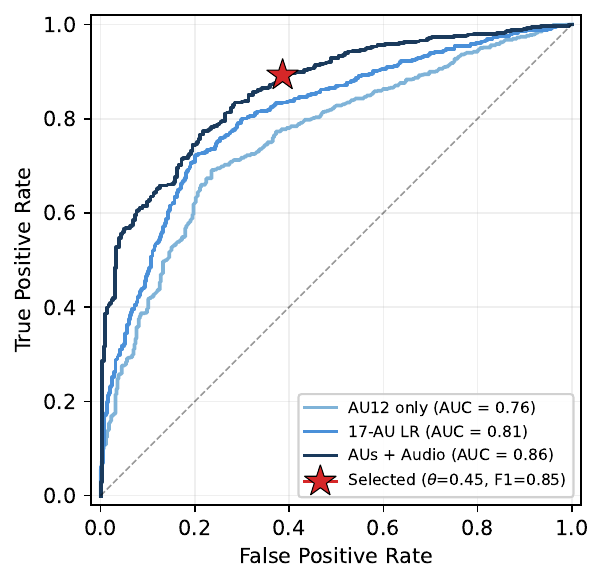}
  \caption{AUC curves for three smile
    detectors. The star marks the F1-peak
    ($\theta{=}0.455$; F1\,=\,0.85, precision\,=\,0.81,
    recall\,=\,0.89) for the final smile detector.}
  \Description{ROC curves comparing three logistic regression smile detectors, with a star marking the F1-maximizing operating point of the final model.}
  \label{fig:roc}
\end{figure}

\begin{figure*}[t]
  \centering
  \includegraphics[width=0.9\textwidth]{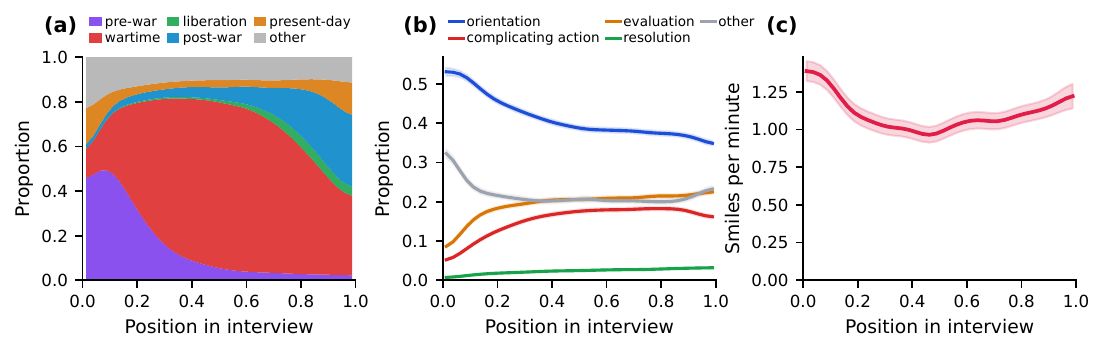}
  \caption{Temporal arc of LLM-annotated narrative features and smile rate averaged across all subjects. \textbf{(a)}~Narrative era proportions. \textbf{(b)}~Narrative structure proportions. \textbf{(c)}~Mean smile rate (smiles/min) with shaded standard deviation.}
  \Description{Three-panel figure showing how narrative era, narrative structure, and mean smile rate change over the course of interviews.}
  \label{fig:temporal-arc}
\end{figure*}

\paragraph{Additional features.} In addition to the initial AU features occurring during the smile,  we calculated per-subject z-scores of AUs and AU values preceding and occurring after the smile itself and VAD affect scores from two new modalities for all timeframes, via audio and via eyegaze, for a total of 150 features.

\paragraph{Model comparison.} We present the AUC of three logistic regression models for binary smile detection trained on different features in Figure~\ref{fig:roc} with a 20-fold cross-validation setup:
\begin{enumerate}
    \item AU12 during smile
    \item 17 AUs occurring during smile
    \item z-score AUs + raw AUs + audio (before/during/after smile)
\end{enumerate}

\paragraph{Feature analysis.} Univariate experiments reveal that the most predictive features were the per-subject z-scores of AU06 (cheek raising, associated with the Duchenne (i.e., genuine) smile \cite{duchenne_1862, ekman_1990}) and AU12 (lip corner pulling), both alone achieving AUCs of 0.809 and 0.756, respectively. Both indicate that \textbf{difference in AU intensity is more predictive than raw AU intensity} for smile prediction, consistent with existing facial Action Unit literature \cite{bartlett_2006, girard_2015, norface_2024}.

For multimodal interaction, we found that emotion derived from audio consistently improved AUC (+0.03 across five configurations) while eyegaze emotion had little effect (+0.002 across five configurations). We hypothesize this asymmetry is a result of AU signal largely overlapping with eyegaze information while audio is mostly independent from observable physical features.

\paragraph{Final smile detection.}
We run the best model (AUs + Audio) on the entire 3,997 video set using the F1-maximizing $\theta{=}0.455$ threshold (F1=0.85; precision\,=\,0.81,
recall\,=\,0.89). This model narrowed the original set of 287,991 smile candidates to 126,812 selected smiles, which forms the main smile corpus.

\section{Narrative Feature Processing}
\label{sec:open-coding}

In addition to pre-existing data on valence derived from audio \cite{lertpetchpun_2025, ma_2026} and eye gaze \cite{ma_2026}, we run a large-scale automatic transcript labeling task via LLM inference to derive additional narrative features and another human annotation task to understand valence distribution.

\begin{figure*}[t]
  \centering
  \includegraphics[width=0.9\textwidth]{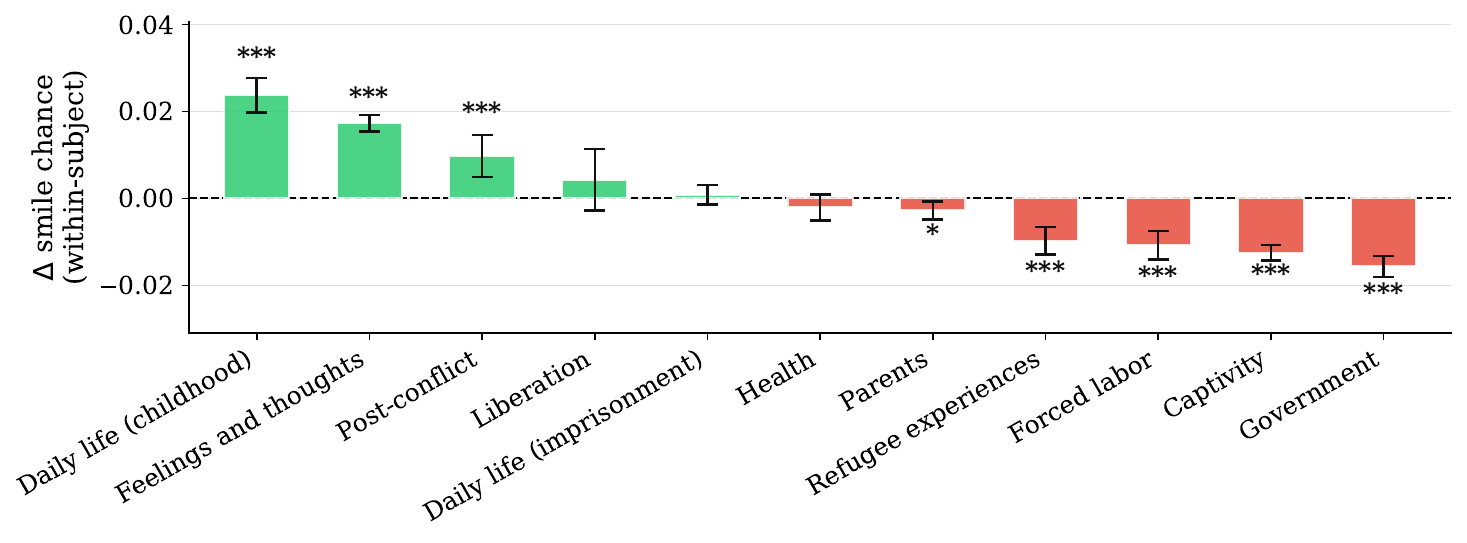}
  \caption{
    Within-subject change in smile presence ($\Delta$ smile presence) for various transcript topics. Rate changes are in relation to the per-speaker smile rate baseline.
  }
  \Description{Horizontal bar chart of within-subject changes in smile presence across transcript topics relative to each speaker baseline.}
  \label{fig:smile-topic}
\end{figure*}

\subsection{Large-Scale Transcript Annotation}
\label{subsec:llm_annotation}

For every sentence in each video, we determine the following seven features from the interview transcript:

\begin{enumerate}
    \item \textbf{Narrative era:} whether the sentence occurred while discussing pre-war events, wartime or camp events, liberation, post-war life, present-day life, or other.
    \item \textbf{Temporal syntax:} the tense of narration, including the strict past (``this happened''), habitual past (``this was happening''), present-tense narration (``this is happening''), and present-tense reflection (``today, I realize''). Present-tense narration has strong ties to PTSD and trauma \cite{herman_1997} and dramatic narrative emphasis \cite{romaine_1984}.
    \item \textbf{Oral structural narrative:} a simplified version of \citeauthor{labov_1967}'s oral narrative taxonomy \cite{labov_1967}, where narrative events fall into ``orientation/scene-setting,'' ``complicating action,'' ``evaluation,'' ``resolution/coda,'' and ``other.'' There has been substantial prior work demonstrating the effectiveness of automatic NLP methods for this task \cite{levi_2022, saldias_2020} including on Holocaust testimonies \cite{shizgal_2025}.
    \item \textbf{Topics:} whether the sentence is related to any of the topics of ``Parents,'' ``Captivity,'' ``Daily life (childhood),'' ``Daily life (imprisonment),'' ``Feelings and thoughts,'' ``Forced labor,'' ``Government,'' ``Health,'' ``Liberation,'' ``Post-conflict,'' or ``Refugee experiences.''
    \item \textbf{Memory recall type:} whether memory recall was semantic or episodic \cite{levine_2002, conway_2005}, using \citeauthor{levine_2002}'s Autobiographical Interview terminology of ``external'' (describing places, things, habits) or ``internal'' (describing a specific event in time).
    \item \textbf{Valence of the narrative:} the inferred emotion of how the subject felt during the narrative based solely from described events, ignoring present-day commentary.
    \item \textbf{Present-day valence:} the inferred emotion of the subject as they are speaking, inferred based on how they interact with and describe the narrative content.
\end{enumerate}

\begin{figure*}[t]
  \centering
  \includegraphics[width=0.8\textwidth]{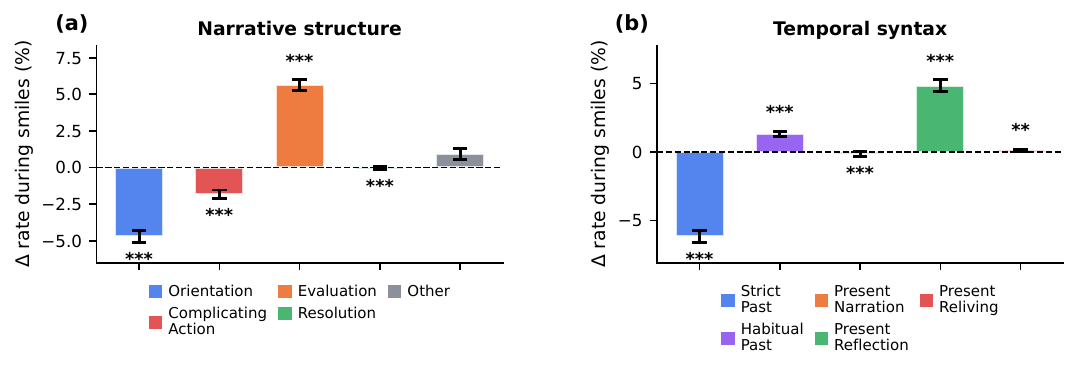}
  \caption{Change in rate of \textbf{(a)} narrative structure and \textbf{(b)} temporal syntax categories during smile vs. non-smile sentences.}
  \Description{Two-panel bar chart comparing rates of narrative structure and temporal syntax categories in smile versus non-smile sentences.}
  \label{fig:narrative-memory}
\end{figure*}

We process transcripts via inference of the open-source LLM GPT-OSS-120B, providing sentences starting up to 20 seconds before and 15 seconds after the smile as context. The first five involve syntactic and semantic classification, tasks for which LLMs have demonstrated sensitivity to linguistic structure and grammatical constructions\cite{tenney_2019, hu_2024}, while the latter two are inferences on speakers' emotions in the past and in the present. LLMs show promise in sentiment analysis tasks during oral narratives \cite{cherukuri_2025} but also fail to capture nuances of human subjectivity \cite{ma_2025, santurkar_2023}. We recognize this limitation and use LLM annotations in this work as a logistical means to annotate narrative roles at scale. In preliminary checks, minor prompt and model changes had little effect on final label distributions with variability analogous to variation among human annotators. We annotated 1{,}083{,}417 sentences across the entire video corpus. We visualize how three features evolve over the course of the interview in Figure~\ref{fig:temporal-arc}. Both narrative era and narrative structure follow predictable patterns; we observe smile rate peaks during interview beginnings, reaches its nadir in the middle of the interview, which coincides with the highest proportion of wartime content, and increases again as the interview shifts to post-war and present-day discussion. Smile rate is lowest when wartime content peaks and rises again in post-war and present-day segments.

\subsection{Valence Annotation}
\label{subsec:valence_annotation}

To validate the LLM valence labels, four annotators independently labeled the same 60 smiles as positive, neutral, or negative for both narrative valence and present-day valence in the same interface as Figure~\ref{fig:ui}. We find the inter-annotator agreement similarly low to the pilot (\S\ref{subsec:low_agreement}): Fleiss' $\kappa{=}0.24$ for narrative valence and $\kappa{=}0.30$ for present-day valence \cite{fleiss_1971}, with unanimous agreement on only 20\% and 31\% of tasks, respectively. Compared to human-human agreement, we find LLM-human agreement is actually slightly \textbf{better} for narrative valence $\kappa{=}0.26$ but much worse for present-day valence $\kappa{=}0.09$, as a text-only transcript does not contain body language and speech patterns highly associated with emotional state \cite{zeng_2009}.

\begin{table}[t]
\centering
\caption{Accuracy (macro F1) of each modality against human majority-vote valence labels.}
\label{tab:valence-alignment}
\begin{tabular}{lccc}
\toprule
& \textbf{Transcript} & \textbf{Audio} & \textbf{Eyegaze} \\
\midrule
\textbf{Narrative}   & \textbf{52.5\% (0.49)} & 44.7\% (0.30) & 42.4\% (0.29) \\
\textbf{Present-day} & 45.8\% (0.33) & \textbf{72.3\% (0.60)} & 67.8\% (0.58) \\
\bottomrule
\end{tabular}

\end{table}

\paragraph{Modality-valence alignment.} We determine which of the three modalities (transcript, audio, eyegaze) best aligns with the two emotion types (narrative and present-day) to pick label modalities for corpus-scale smile analysis. We use transcript labels as-is and discretize audio and eye emotion into positive, neutral, and negative labels then compare against human majority-vote labels in Table~\ref{tab:valence-alignment}. We find narrative valence is best predicted by LLM transcript outputs (53\%) while present-day valence is best predicted by audio emotion (72\%), and opt to use these two modalities as respective valence labels.

\section{Smiling, Emotion, and Narrative Patterns}
\label{sec:narrative-changes}

In four analyses, we address how the presence and type of smile affects the emotions and narratives of the survivors. In \S\ref{subsec:topic-smiling}, we see which semantic topics are most associated with smiling; in \S\ref{subsec:narrative-memory}, we find which narrative structures and temporal syntaxes most occur with smiling; in \S\ref{subsec:valence-trajectories} we observe how negative-affect smiles improve valence trajectories across all modalities and both emotion types; and in \S\ref{subsec:gaze-dynamics} we see how smiling changes eye dynamics and blink rates.

\subsection{Topics Correlate with Smile Rates}
\label{subsec:topic-smiling}

We aim to understand which topics labeled from \S\ref{subsec:llm_annotation} occur most often with smiling. We define smile activation probability for each word as the chance a smile occurred within 0.5 seconds of word onset, then offset with a per-subject baseline smile rate to get probability deltas. We combine similar words such as ``running'' and ``ran'' into the same word lemma of ``run,'' then sum per-lemma deltas and average by topic. We plot smile frequencies in Figure~\ref{fig:smile-topic}. Intuitively, topics with typically positive valence (childhood, liberation) see significantly higher smile rates while negative topics (captivity, forced labor) see suppressed smile rates.

\subsection{Smiling Is Associated with Narrative Structure and Temporal Syntax}
\label{subsec:narrative-memory}

We compare per-subject rates of each narrative-structure and temporal-syntax class in sentences with vs.\ without a nearby smile in Figure~\ref{fig:narrative-memory} via paired Wilcoxon tests. We observe two related patterns: first, smiling significantly reduces the rates of Orientation (-4.70\%) narrative sentences and sentences set in the Strict Past (-6.15\%), and significantly increases the rates of Evaluation narrative sentences (+5.62\%) and sentences in the Present Reflection (+4.84\%). We hypothesize both phenomena reflect shifts between two patterns of interviewee discourse; during matter-of-fact narration, tied with orientation and the strict past, the subject withholds emotional processing and interviewer socialization, and thus smiling, and during present-day reflection, the subject provides commentary and social interaction with the interviewer, which engages more emotional and social behaviors including smiling. This is consistent with literature finding that social function modulates smiling more so than pure emotional expression \cite{fridlund_1994} and that nonverbal expressions of emotion such as smiling during narration are timed more for communicative function than narrative fidelity \cite{bavelas_2000}. Our findings contribute to the overall consensus that smiling itself serves a large role in social interaction and communication \cite{chovil_1991, fridlund_1994, ekman_1990, martin_2017} beyond just individual emotional expression.

\begin{figure}[t]
  \centering
  \includegraphics[width=\columnwidth]{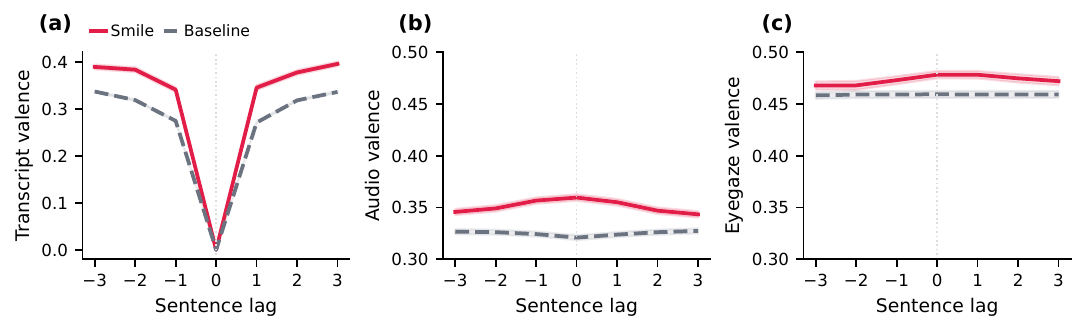}
  \caption{Valence trajectories when \textit{narrative valence via transcript} is negative for smiling events (solid red) versus non-smiling events (dashed gray).}
  \Description{Line plot of narrative valence trajectories around negative-valence smiles versus matched non-smile events.}
  \label{fig:valence-trajectories}
\end{figure}

\begin{figure}[t]
  \centering
  \includegraphics[width=\columnwidth]{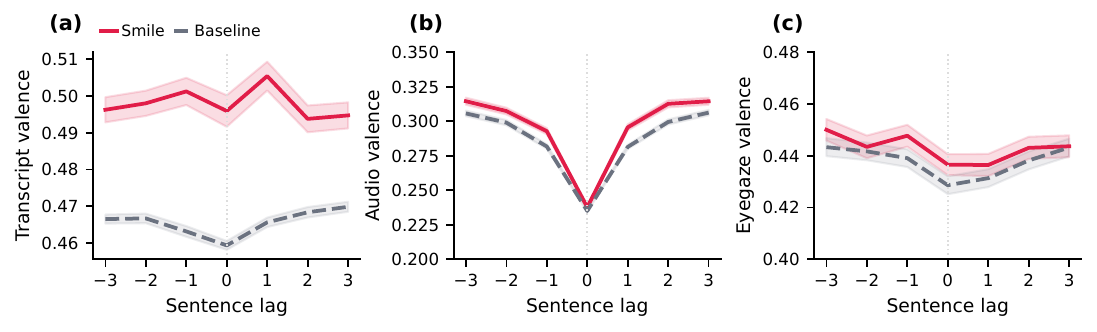}
  \caption{Valence trajectories when \textit{present-day valence via audio} is negative for smiling (solid red) versus non-smiling (dashed gray).}
  \Description{Line plot of present-day valence trajectories around negative present-day-valence smiles versus matched non-smile events.}
  \label{fig:pd-valence-trajectories}
\end{figure}

\begin{table}[t]
\centering
\caption{Valence trajectories at $t{=}+1$ between negative-valence smile clusters and no-smile clusters, with paired Wilcoxon test $p$-value on within-subject means, Cohen's $d$ effect sizes, and \% of subjects with higher valence.}
\label{tab:smile-valence}
\small
\begin{tabular}{@{}lcccccc@{}}
\toprule
& \multicolumn{3}{c}{\textbf{Narrative valence}} & \multicolumn{3}{c@{}}{\textbf{Present-day valence}} \\
\cmidrule(lr){2-4} \cmidrule(l){5-7}
\textbf{Modality} & $d$ & \% Higher & $p$ & $d$ & \% Higher & $p$ \\
\midrule
Transcript & $0.50$ & 77\% & $1.3{\times}10^{-50}$ & $0.36$ & 74\% & $2.7{\times}10^{-42}$ \\
Audio      & $0.55$ & 75\% & $5.6{\times}10^{-56}$ & $0.42$ & 72\% & $1.2{\times}10^{-36}$ \\
Eyegaze    & $0.22$ & 61\% & $1.7{\times}10^{-11}$ & $0.06$ & 54\% & $8.5{\times}10^{-3}$ \\
\bottomrule
\end{tabular}
\end{table}

\subsection{Negative-Affect Smiles Accompany More Positive Valence Trajectories}
\label{subsec:valence-trajectories}

We analyze how smiles annotated with \textbf{negative narrative valence} ($n{=}23{,}624$, 18\% of smiles) and \textbf{negative present-day valence} ($n{=}18{,}872$, 19\% of smiles), derived from transcript and audio, respectively, relate to the valence of surrounding sentences. To do so, we map each negative valence smile onto the sentence that overlaps with the smile timestamp and collect the three preceding and following sentences, forming seven-sentence clusters from stimulus onset $t\in [-3, 3]$. For each valence type, we contrast the valence trajectories of these smile sentence clusters to sentence clusters where no smiling occurs ($n=221{,}347$ and $n=227{,}907$) along our three modalities of emotion (transcript, audio \cite{lertpetchpun_2025}, and eyegaze \cite{ma_2026}) for both narrative (Figure~\ref{fig:valence-trajectories}) and present-day (Figure~\ref{fig:pd-valence-trajectories}) valence. Under all six conditions, valence trajectories are higher during smile clusters. We note valence differences at $t=0$ for transcript and audio by definition are $0$ as this was the filter criteria. We report effect sizes, subject coverage, and $p$-values of valence trajectories in Table~\ref{tab:smile-valence}, finding effects are both stronger and more consistent for narrative affect than present-day affect and that transcript and audio are affected more saliently than eyegaze.

\begin{figure}[t]
  \centering
  \includegraphics[width=0.9\columnwidth]{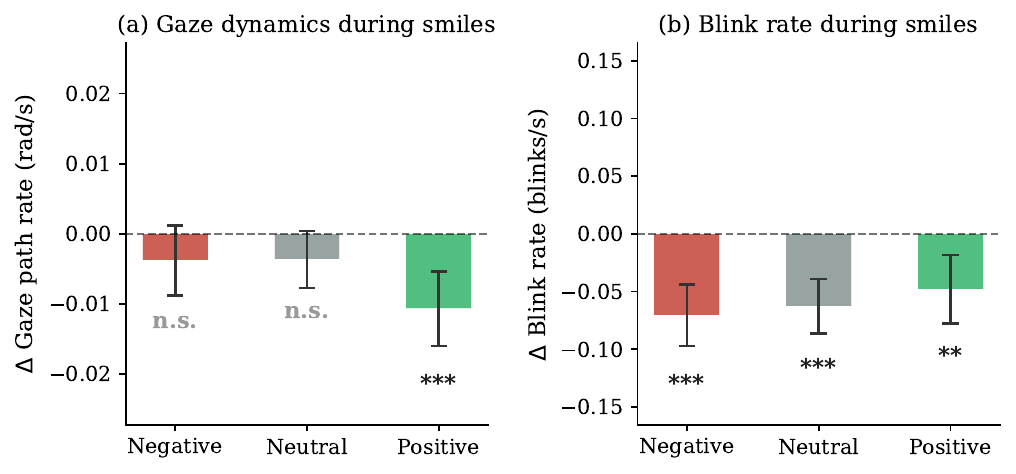}
  \caption{
    Change in \textbf{(a)} gaze dynamics and \textbf{(b)} blink rate during smiling, separated by narrative valence. Values are calculated per-subject and averaged, with 95\,\% CI error bars.
  }
  \Description{Two-panel bar chart of changes in gaze dynamics and blink rate during smiling, stratified by narrative valence with confidence intervals.}
  \label{fig:gaze-blink}
\end{figure}

\subsection{Smiling Is Associated with Lower Eye Gaze Dynamics and Blink Rates}
\label{subsec:gaze-dynamics}

We examine how eye gaze movements are influenced during smiling. While fine-grained eye movements such as fixation and saccades are well-studied and associated with cognitive processing \cite{rayner_1998} and social interaction \cite{kleinke_1986}, the low video and temporal resolution (typically 320x240pp and 30 fps) and unknown interviewer location prevented us from directly modeling established behaviors such as direct eye contact, gaze aversion, and ``staring off into space'' (which itself is well known to be associated with dissociation and memory recall \cite{glenberg_1998}). However, we find that two coarse eye movement features, total gaze dynamics and blink rate, have strong deviations from baseline during specific types of narrative-valence smiles. We find no significant effects running the same experiments stratified on present-day valence, highlighting the established role of eye gaze in memory and narrative recollection \cite{richardson_2000, johansson_2014}.

\paragraph{Gaze dynamics.}
We measure gaze dynamics as the total angular distance the eyes move over time, which we calculate as radian displacement per second (rad/s) averaged across both eyes. We again separate smiles by their associated narrative valence, and find that while \textit{neutral} and \textit{negative} smiles have gaze dynamics on par with the average non-smiling moment, \textit{positive} smiles have significantly reduced eye movement ($\Delta{=}-0.01$, $p{<}0.001$); cf. Figure~\ref{fig:gaze-blink} (a). This is qualitatively consistent both with the open coding from \S\ref{sec:open-coding}, where happy smiles seemed to be associated with direct interviewer eye contact and/or fond memory recall with a fixed gaze during recollection; and with existing eye movement literature finding that eye dynamics decrease during sustained social interactions \cite{kleinke_1986} and when averted during memory recall \cite{glenberg_1998}.

\paragraph{Blink rate.}

We have direct frame-level binary blink labels from AU45 from the OpenFace features and calculate blink rate as blinks per second. We find blink rate is lower during all smiles, but observe that as narrative content becomes more negative, it drops more ($\Delta{=}-0.07$, $p<2.5\text{e}{-}7$ for negative vs. $\Delta{=}-0.05$, $p<1.5\text{e}{-}3$ for positive); cf. Figure~\ref{fig:gaze-blink} (b). The relationship between blinking and cognitive engagement is well-studied, with dual evidence finding blink rate increases when the mind wanders \cite{Smilek_2010} and is suppressed during increased cognitive load \cite{holland_1972, stern_1984}. One interpretation of our findings is that smiling utilizes higher cognitive engagement, explaining the suppressed blinking, and that negative-affect smiling in particular is more effortful than positive smiling. Negative-valence smiling may function as a form of emotional regulation \cite{ekman_1982}, which has long been linked to increased cognitive effort \cite{richards_1999, richards_2000}.

\section{Cultural Differences and Limitations}
\label{sec:cultural}
Smiling is both an instinctual and learned behavior \cite{freedman_1964, ekman_1969}. It is important to consider the role that cultural influences play in interpersonal interaction and facial expressions. Our findings are limited by the cultural backgrounds of the subjects, and for example, we highlight one particular survivor from late Tsarist Russia and the Soviet Union who emigrated to the USA after the war. The survivor does not smile and describes her wartime experiences in a virtually emotionless manner. Taken in isolation and scrutinized via our analysis, this would imply a lack of felt emotion from the subject; however, in the context of the interview, it is revealed the survivor was socialized in a society where, due to the terror directed by the Soviet state, it could be a matter of survival to not show any emotions. While smiling typically demonstrates politeness and appreciation, this communication in Russian-speaking regions is often expressed through seriousness \cite{sheldon_2017, sternin_2000}. This cultural trait can still be observed today in Russian-speaking regions and differs significantly from cultural norms in Western societies \cite{arapova20165672}. The social and narrative role of smiling is heavily implicated by the personal context and smilers' identities.

\section{Safe and Responsible Innovation Statement}

In this work, we study Holocaust survivor video testimonies through facial, voice, and narrative processing. Analysis of this atrocity carries significant caution and ethical weight. All video was provided by the USC Shoah Foundation's Visual History Archive (VHA) under institutional agreement, with no new data collection. We do not disclose any of the survivors' names or personally identifiable information but we anonymously acknowledge their contribution and honor the memory of those who did not survive. This work is basic research understanding the role smiles play in traumatic narratives, and as a result, we report only corpus-level and anonymous statistics. We must consider the bias of cultural influence on smiling as well. We used generative AI to assist with coding, references, and building the annotation tool.

\section{Conclusion}
\label{sec:conclusion}

We study the role smiling plays for regulating the emotional and social dynamics of narrative recollection and discover the multimodal interactions of smiling with valence, physical markers, and narrative content. These findings underpin how smiling is an incredibly complex and rich expression of human communication and hope this work inspires future research on human smile.

\begin{acks}
This work was supported by the USC Shoah Foundation, the Luxembourg National Research Fund (Grant C24/ID/18896236/VOICES), the German Research Foundation (551119792), the Andrew and Erna Viterbi Graduate Student Fellowship, and the Charles Lee Powell Foundation.
\end{acks}

\bibliographystyle{ACM-Reference-Format}
\bibliography{references}

\end{document}